\documentclass[11pt,fleqn]{article}       
\usepackage{geometry}
\usepackage[T1]{fontenc}
\usepackage[utf8]{inputenc}
\usepackage{authblk}
\usepackage{mathrsfs}
\usepackage{scalerel}
\usepackage[font=small,labelfont=bf]{caption}
\usepackage{graphicx}
\usepackage{multirow}
\usepackage{amsmath,amssymb,amsfonts}
\usepackage{verbatim}
\usepackage{bm}
\usepackage{color}
\usepackage{ulem}
\usepackage{mathtools}
\usepackage{cite}
\usepackage{colortbl}
\definecolor{light-gray}{gray}{0.85}
\usepackage[pdftex,colorlinks,pdfpagelabels]{hyperref}
\hypersetup{
   bookmarksnumbered,
   citecolor={blue},
   linkcolor={blue},
   urlcolor={blue},
   filecolor={blue}
} 
\newcommand{\eVdist}{\kern-0.06em}

\newcommand{\mev}{\:\text{Me\eVdist V}}

\newcommand{\re}{\:\text{Re}\,}
\newcommand{\im}{\:\text{Im}\,}
\newcommand{\tev}{\:\text{Te\eVdist V}}

\newcommand{\kIR}{\ensuremath{\kappa_{\scaleto{1,\text{R}}{6pt}}}}
\newcommand{\kIL}{\ensuremath{\kappa_{\scaleto{1,\text{L}}{6pt}}}}
\newcommand{\kIIR}{\ensuremath{\kappa_{\scaleto{2,\text{R}}{6pt}}}}
\newcommand{\kIIL}{\ensuremath{\kappa_{\scaleto{2,\text{L}}{6pt}}}}
\newcommand{\AddrMichigan}{%
\textit{Leinweber Center for Theoretical Physics, Department of Physics, University of Michigan, Ann Arbor, MI 48109, USA}
}
\newcommand{\AddrStockholm}{
\textit{The Oskar Klein Centre for Cosmoparticle Physics, Department of Physics, Stockholm University, Alba Nova, 10691 Stockholm, Sweden}
}

\usepackage{fancyhdr}
 
\fancypagestyle{plain}{%
    \fancyhead[R]{LCTP-19-22}
    
}
\date{}
\title{\Large\bf Baryogenesis from a Modulus Dominated Universe}
\author[1]{Gordon Kane}
\author[2]{Martin Wolfgang Winkler\thanks{martin.winkler@su.se}}
\affil[1]{\AddrMichigan}
\affil[2]{\AddrStockholm}

\begin{document}
\maketitle
\vspace*{0mm}
\begin{abstract}
String/ M-theory compactifications predict the existence of a modulus field with a mass of $100-10000\tev$. Its decay at MeV-temperatures generates large amounts of entropy and washes out any previously produced baryon asymmetry. We describe how the baryon asymmetry can be (re)generated by the modulus decay. The mechanism relates the smallness of the asymmetry to the hierarchy between the Planck- and the Fermi-scale.
\end{abstract}
\clearpage

\section{Introduction}
Understanding the matter asymmetry of the universe is one of the most challenging problems in particle physics and cosmology.  At the end of inflation the universe was an unstable energy density that decayed into matter and anti-matter, presumably in equal amounts. Today the number of anti-baryons is about 1 ten-billionth the number of baryons. The problem is not that we cannot imagine how a matter asymmetry came about, but that there are too many proposals. In most of them, the magnitude of the asymmetry is a mere accident and relies on adhoc choices of free parameters.

In this work, we will relate the smallness of the baryon asymmetry to the hierarchy between the Fermi and the Planck scale. A wide class of string/ M-theory compactifications share the presence of a relatively light modulus field $\varrho$ with only gravitational couplings. We will show that its mass $m_{\varrho}\simeq 10^2-10^4\tev$ is not arbitrary, but is fixed by the strength of supersymmetry breaking which determines the mass of the Standard Model Higgs boson. The modulus gets displaced during inflation and dominates the energy content of the universe when it decays at temperature $T\simeq 10-100\mev$, shortly before nucleosynthesis~\cite{Coughlan:1983ci}. This causes a second reheating and washes out any previously produced baryon asymmetry. Leptogenesis or electroweak baryogenesis are doomed to fail in such a universe.
 
However, we will demonstrate that the modulus decay chain itself successfully generates the matter asymmetry.\footnote{See~\cite{Kitano:2008tk,Allahverdi:2010im,Allahverdi:2010rh,Ishiwata:2013waa,Dhuria:2015xua} for complementary attempts to generate the baryon asymmetry by the decay of a modulus field.} The mechanism consists of two steps: first the modulus decays to gauginos with a branching ratio of $\mathcal{O}(1)$. Then, baryons are created in the gaugino decay. The necessary CP violation originates from the phase of the gaugino mass matrix, whereas baryon number violation occurs through the R-parity violating three-quark operator $W\supset UDD$.
While baryogenesis by gaugino decay has been considered previously~\cite{Cui:2013bta,Arcadi:2013jza,Arcadi:2015ffa}, this was done in the context of a thermal universe. The thermal history, however, proves challenging since washout processes (reverse reactions) tend to dilute any previously created baryon asymmetry. The wash-out problem completely disappears in the non-thermal framework, where the baryons are created from modulus decay. This is because the decay temperature is far too cold for wash-out processes to play any role. 

We determine the matter asymmetry in the Minimal Supersymmetric Standard Model (MSSM) and in its extension by a hidden sector gaugino (following the approach of~\cite{Pierce:2019ozl}). The final baryon abundance is essentially determined by $\sqrt{m_{\varrho}/M_\text{P}}$ times a loop factor, with the modulus mass tied to the Fermi scale as argued above.

\section{Modulus Cosmology}

Supergravity/ string theories generically feature light uncharged scalar fields with flat potentials (moduli). The lightest modulus $\varrho$ gets displaced from its vacuum expectation value $\varrho_0$ during inflation. This can easily be seen by expanding the supergravity scalar potential around $\varrho_0$. Inflation induces a linear term of the form 
\begin{equation}
  V\supset \left(3H^2\,K_\varrho\big|_{\varrho_0}\right)\,\varrho\,,
\end{equation}
where $H$ denotes the Hubble scale during inflation. The derivative of the K\"ahler potential $K_\varrho$ is generically non-vanishing for a modulus and drives $\varrho$ away from $\varrho_0$. Additional displacement may arise from thermal effects~\cite{Buchmuller:2004xr}.

After reheating, the universe is radiation-dominated with a subdominant energy fraction stored in the modulus degree of freedom. The postinflationary evolution of $\varrho$ is determined by its equation of motion
\begin{equation}
  \overset{..}{\varrho} + 3H\,\overset{.}{\varrho} + m_{\varrho}^2\,\varrho = 0\,,
\end{equation}
for canonically normalized $\varrho$. The modulus remains fixed as long as $H>m_{\varrho}$ and then performs coherent oscillations around its minimum. In this period, the modulus energy density redshifts as $\rho_{\varrho}\propto T^3$ with $T$ denoting the temperature of the thermal bath. Since the radiation energy density decreases as $\rho_{\gamma}\propto T^4$, the modulus contribution becomes more significant as the universe cools down. After some time, the modulus typically dominates the energy content of the universe~\cite{Coughlan:1983ci}. When it decays, it reheats the universe a second time. But since the decay rate is Planck-suppressed, this happens only shortly before primordial nucleosynthesis. The corresponding decay temperature is 
\begin{equation}
 T_\varrho = \left(\frac{90}{\pi^2 g_*}\right)^{1/4} \sqrt{M_\text{P}\Gamma_\varrho}\,\simeq\, 20\mev \,\sqrt{c}\, \left(\frac{m_\varrho}{100\tev}\right)^{3/2}\,,
\end{equation}
where we have expressed the modulus decay rate as
\begin{equation}
 \Gamma_\varrho = c\, \frac{m_\varrho^3}{M_\text{P}^2}\,.
\end{equation}
The constant $c$ is typically of $\mathcal{O}(1)$. Hence, the modulus decay only reheats the universe to temperatures of $\text{MeV}-100\mev$.

While successful nucleosynthesis can be realized if $T_\varrho\gtrsim 5\mev$~\cite{deSalas:2015glj,Hasegawa:2019jsa}, popular baryogenesis mechanisms (e.g.\ leptogenesis, electroweak baryogenesis) operate at much higher temperatures. Therefore, we are forced to look for an alternative.

In order to identify a suitable low temperature baryogenesis scheme, it is instructive to look at the modulus abundance $Y_\varrho$ prior to decay,
\begin{equation}\label{eq:yield}
 Y_\varrho = \frac{3}{4} \left(\frac{90}{\pi^2\,g_*}\right)^{1/4}\sqrt{\frac{c\,m_\varrho}{M_P}}\,\simeq\, 1.5\times 10^{-7}\,\sqrt{\frac{c\;m_\varrho}{100\tev}}\,.
\end{equation}
The number of $10^{-7}$ multiplied by a loop factor is intriguingly close to the observed baryon asymmetry $Y_b=0.8\times 10^{-10}$. A very natural possibility is, hence, to consider the modulus decay chain as the origin of baryons. Intriguingly, the smallness of the baryon asymmetry would then be explained by the hierarchy between the Planck scale and the supersymmetry breaking scale. The latter sets the modulus mass and is again strongly correlated with the electroweak scale.

Before we describe the baryogenesis mechanism in detail, it is important to obtain a more precise estimate of the modulus abundance. For this purpose, we will now derive the modulus decay rate and the modulus mass in prominent UV theories.

\section{Modulus Decay}

We start by considering a generic supergravity theory with superpotential $W$, K\"ahler potential $K$ and gauge kinetic function $f$ for the standard model gauge groups. The modulus $\varrho$ may appear in $f$ and we take $\re \varrho$ to be the cosmologically relevant field which decays. Following the standard convention, we define the $F$-terms
\begin{equation}
  F^{m}= e^{G/2} K^{m\bar{n}} G_{\bar{n}}\,,
\end{equation}
where the index $\bar{n}$ indicates differentiation with respect to the field $\bar{\varphi}_n$, while $K^{m\bar{n}}$ denotes the inverse K\"ahler metric. Furthermore, we introduced $G= K + \log|W|^2$. We set $M_{\text{P}}=1$ in the following, but recover it in the final expressions of decay rates.

Let us first turn to the modulus interactions with gauge fields which derive from the Lagrangian~\cite{Brignole:1997dp}
\begin{equation}
  \mathcal{L}_{\text{gauge}}\supset -\frac{1}{4} \re\!f \;F^a_{\mu\nu} F^{a\,\mu\nu} + \frac{i}{2} \re\!f \; \left(\bar{\lambda}^a \bar{\sigma}^\mu\partial_\mu \lambda^a+\text{h.c.}\right) -\left(\frac{F^m \partial_m f}{4} \lambda^a\lambda^a + \text{h.c.}\right)\,.
\end{equation}
The first, second and third term are the gauge boson, gaugino kinetic terms and gaugino soft mass terms \footnote{We neglected anomaly-mediated contributions which hardly affect the modulus decay rate into gauginos.} respectively. The index $m$ runs over all hidden sector fields including the modulus. The corresponding modulus decay rates into gauge bosons and gauginos are (cf.~\cite{Nakamura:2006uc})\footnote{We assume a linear dependence of $f$ on the hidden sector fields such that higher derivatives of the gauge kinetic function do not contribute.}
\begin{align}
  \Gamma_{gg} &= \frac{N_{g}\,m_\varrho^3}{128\pi}\,\frac{|\partial_\varrho f|^2}{K_{\bar{\varrho}\varrho}\,(\re f)^2}\,,\label{eq:Gammagg}\\
  \Gamma_{\tilde{\lambda}\tilde{\lambda}} &=
  \frac{N_{g}\,m_\varrho }{128\pi}\,\frac{\left|(F^m_{\;\;\varrho}+F^m_{\;\;\bar{\varrho}})\,\partial_m f\right|^2}{K_{\bar{\varrho}\varrho}\,(\re f)^2}\,,\label{eq:Gammagaga}
\end{align}
where we neglected the mass of the final state particles. The number of gauge bosons (= number of gauginos) $N_g=12$ in the MSSM. Notice that in the limit of a heavy supersymmetric modulus $F^m_{\;\;\bar{\varrho}}\simeq F^\varrho_{\;\;\bar{\varrho}} \;\delta_{m\varrho}$ and
\begin{equation}\label{eq:heavymodulus}
  F^\varrho_{\;\;\bar{\varrho}}\simeq e^{G/2} G_{\bar{\varrho}\bar{\varrho}} \,K^{\varrho\bar{\varrho}} \simeq m_\varrho\,,
\end{equation}
which implies $\Gamma_{\tilde{\lambda}\tilde{\lambda}}\simeq \Gamma_{gg}$.

We next consider Higgs and higgsino final states. For simplicity, the K\"ahler metric of the Higgs fields is taken to be independent of the modulus. It may otherwise carry a generic hidden sector dependence,
\begin{equation}
  K= \hat K(\varrho,\bar{\varrho},\varphi_m,\bar{\varphi}_m) + \hat{Z}(\varphi_m,\bar{\varphi}_m) \,(\bar{h}_u h_u+ \bar{h}_d h_d )+ \left[Z(\varphi_m,\bar{\varphi}_m) \,h_d h_u + \text{h.c.}\right]\,.
\end{equation} 
The modulus decay rates into Higgs bosons and higgsinos (summed over all up- and down-type states) are then determined as\footnote{We assume absence of any direct modulus-Higgs couplings in the superpotential.}
\begin{align}
  \Gamma_{HH} &= \frac{\left|(\partial_\varrho+\partial_{\bar{\varrho}})\,m_{H}^2\right|^2+\left|(\partial_\varrho+\partial_{\bar{\varrho}})\,B\mu\right|^2}{8\pi m_\varrho\,K_{\varrho\bar{\varrho}}}\,,\label{eq:GammaHH}\\
  \Gamma_{\tilde{h}\tilde{h}} &= \frac{m_\varrho \left|(\partial_\varrho +\partial_{\bar{\varrho}})\,\mu \right|^2}{8\pi\,K_{\varrho\bar{\varrho}}}\,,\label{eq:Gammahihi}
\end{align}
where we again neglected any final state phase space suppression. The mass parameters in the above expressions are defined as~\cite{Brignole:1997dp}
\begin{align}
\mu&=\frac{1}{\hat Z}\left(m_{3/2}\,Z-F^{\bar{m}} \partial_{\bar{m}} Z\right)\,,\\
m_H^2 &= |\mu|^2 + m_{3/2}^2 - F^{\bar{m}} F^n\partial_{\bar{m}}\partial_n \log \hat Z\,,
\label{eq:mH}\\
B\mu&= \frac{1}{\hat Z}\left[2\,m_{3/2}^2\,Z- m_{3/2} F^{\bar{m}}\partial_{\bar{m}} Z + m_{3/2} F^m \left(\partial_m Z - 2 \,Z\,\partial_m \log\hat Z\right)\right.\nonumber\\
&\phantom{= \frac{1}{\hat Z}}\;\; \left. -F^{\bar{m}} F^n \left(\partial_{\bar{m}}\partial_n Z - 2 \partial_{\bar{m}} Z\,\partial_n \log\hat Z\right)\right]\,.
\end{align}
The decay rate of the modulus to sfermions can be determined from~\eqref{eq:GammaHH} by setting $Z$ to zero and replacing $\hat Z$ with the sfermion K\"ahler metric. Furthermore, if kinematically accessible, the modulus decay to gravitinos $\tilde{\Psi}$ occurs with the rate~\cite{Endo:2006zj}
\begin{equation}\label{eq:Gamma32}
  \Gamma_{\tilde{\Psi}\tilde{\Psi}}=\frac{m_\varrho^5}{288\pi\,m_{3/2}^2}\frac{\left|G_\varrho\right|^2}{K_{\varrho\bar{\varrho}}}\,.
\end{equation}
In some modulus stabilization schemes, the imaginary part $a=\im\varrho$ is a (nearly) massless axion. If such an axion exists, it can get pair-produced by modulus decay. The corresponding rate derives from the axion kinetic term and reads~\cite{Higaki:2013lra}
\begin{equation}\label{eq:axionrate}
  \Gamma_{aa}=\frac{1}{64\pi}\frac{ K_{\varrho\bar{\varrho}\varrho}^2 \,m_\varrho^3}{ K_{\varrho\bar{\varrho}}^3}\,.
\end{equation}

Let us now turn to some concrete ultraviolet theories for which we can derive the modulus decay rate explicitly. As a first example we consider KKLT modulus stabilization in type IIb string theory~\cite{Kachru:2003aw}.\footnote{For moduli decay rates in the Large Volume Scenario see~\cite{Cicoli:2012aq,Hebecker:2014gka}.} The original setup assumes that all complex structure moduli of a compact Calabi Yau manifold and the dilaton are stabilized by fluxes~\cite{Giddings:2001yu}. The low energy effective theory contains a single lightest K\"ahler modulus which parameterizes the volume of the compact manifold. Its K\"ahler potential takes the familiar no-scale form
\begin{equation}
K = -3 \log(T+\bar{T})\,.
\end{equation}
The $T$ modulus itself is stabilized by non-perturbative effects which stem from Euclidean D3 instantons or from gaugino condensation on a stack of D7 branes. A de Sitter vacuum with a small cosmological constant is obtained via uplifting with the $F$-term of a matter field $X$~\cite{Lebedev:2006qq}. While the scalar component of $X$ could also play the role of the lightest modulus, we focus on the case where it decouples from the low energy theory.\footnote{The scalar component of $X$ can be decoupled through Yukawa interactions with heavy fields~\cite{ORaifeartaigh:1975nky,Kallosh:2006dv} or by making $X$ a nilpotent field (see e.g.~\cite{Antoniadis:2014oya,Antoniadis:2014oya,Kallosh:2014wsa}).} Assuming that MSSM gauge fields are living on D7 branes, the gauge kinetic function is given by $f=T$. While $X$ is the dominant source of supersymmetry breaking, the K\"ahler modulus obtains a suppressed but non-vanishing $F$-term~\cite{Choi:2005ge}
\begin{equation}\label{eq:Tfterm}
  F^T\simeq \frac{3\,N}{8\pi^2}\,m_{3/2}\,.
\end{equation}
The stabilization scheme also fixes the mass of $T$ which is dominated by the supersymmetric contribution. One finds\footnote{This relation can be derived from~\eqref{eq:heavymodulus} by setting $W_{TT} = (-8\pi^2/N)\,W_T$ and $W_T \simeq K_T W$ (at leading order).}
\begin{equation}\label{eq:Tmass}
  m_T \simeq \frac{4\pi\,m_{3/2}}{\alpha_{\text{GUT}}\,N}\simeq (1-8)\times 10^3\tev\,,
\end{equation}
with the unified gauge coupling strength $\alpha_{\text{GUT}}\simeq 0.04$. The number $N$ stands for the rank of the hidden sector SU(N) gauge group. In the last step, we employed $N=4-8$ and $m_{3/2}=20 - 100\tev$ as motivated in~\cite{Choi:2005ge}.

The decay rate of the K\"ahler modulus to gauge bosons and gauginos is given by
\begin{equation}\label{eq:Trategauge}
  \Gamma_{gg} \simeq \Gamma_{\tilde{\lambda}\tilde{\lambda}} \simeq \frac{N_{g}\,m_T^3}{96\pi M_{\text{P}}^2}\,.
\end{equation}
In addition, the $T$-dependence of the function $Z$, may induce a large decay rate to Higgs bosons and Higgsinos,
\begin{equation}\label{eq:higgsino}
  \Gamma_{HH}\simeq \Gamma_{\tilde{h}\tilde{h}} \simeq \frac{m_T^3 \left|\partial_T Z\right|^2}{8\pi\,K_{T\bar{T}}{\hat Z }^2}\,,
\end{equation}
where we neglected contributions suppressed by powers of $m_{3/2}/m_T$. We consider the dependence $Z\propto(\bar{T}+T)^{-n}$ with $n=0,\,1/2,\,1$ obtained for matter localized on D3 or D7 branes~\cite{Falkowski:2005ck}.
The decay rate to (s)fermions is chirality-suppressed~\cite{Moroi:1999zb} and can, therefore, be neglected.
Finally, the decay to gravitinos is kinematically accessible. From~\eqref{eq:Tfterm},~\eqref{eq:Tmass} and~\eqref{eq:Gamma32}, we derive 
\begin{equation}
  \Gamma_{\tilde{\Psi}\tilde{\Psi}}=\frac{3}{32\pi}\frac{m_\varrho^3}{M_P^2}\,.
\end{equation}
The gravitinos are themselves long-lived and become non-relativistic before their decay. Since the energy density of non-relativistic matter redshifts slower than the temperature of the thermal bath, gravitinos induce more entropy compared to if they had promptly decayed. We find, however, that gravitinos never dominate the energy content of the universe such that we can neglect the additional entropy release.\footnote{Between the modulus and gravitino decays, the gravitino energy density increases by a factor $1-3$ relative to the energy of the thermal bath. Since, however, only a fraction of the moduli decays into gravitinos, this enhancement does not lead to a gravitino-dominated universe.}

For our second UV example, we turn to orbifold compactifications of the heterotic string. We assume that all geometric moduli are fixed supersymmetrically and can be integrated out. This can potentially be achieved by the interplay between the Fayet–Iliopoulos D-term and non-perturbative physics. The low energy-theory then contains the dilaton as the cosmologically relevant light modulus. Its K\"ahler potential reads
\begin{equation}
  K= -\log(\bar{S}+S)\,,
\end{equation}
while the gauge kinetic function is given as $f=S$. The dilaton can be stabilized by a KKLT-type mechanism which again invokes a hidden sector gaugino condensate~\cite{Lowen:2008fm}. The mass $m_S$ fulfills the same relation~\eqref{eq:Tmass} as the K\"ahler modulus in the type II case. However, for the heterotic string, the hidden sector SU(N) gauge group has to be a subgroup of $E_8$. We will fix $N=4,5$ which is found in the majority of phenomenologically viable orbifold models~\cite{Lebedev:2006tr}. Setting $m_{3/2}\simeq 30-80\tev$ as required to reproduce the observed Higgs mass~\cite{Krippendorf:2012ir,Badziak:2012yg}, we can estimate
\begin{equation}\label{eq:dilatonmass}
  m_S \simeq (2-6)\times 10^3\tev\,.
\end{equation}
The dilaton decay is dominated by gaugino and gauge boson final states, with
\begin{equation}\label{eq:dilatonrate}
  \Gamma_{gg} \simeq \Gamma_{\tilde{\lambda}\tilde{\lambda}} \simeq \frac{N_{g}\,m_S^3}{32\pi M_{\text{P}}^2}\,.
\end{equation}
Notice a factor of 3 difference compared to the type IIb case which follows from the difference of K\"ahler potentials. Since $S$ does not occur in the K\"ahler potential of the visible sector fields\footnote{For the heterotic string, the K\"ahler metrics of the visible sector field are typically functions of the K\"ahler moduli and not of the dilaton.} and since $m_S\gg m_{3/2}$, we can neglect dilaton decays to the chiral superfields of the MSSM. However, we have to consider the decay to gravitinos which occurs at the rate
\begin{equation}
  \Gamma_{\tilde{\Psi}\tilde{\Psi}}=\frac{1}{32\pi}\frac{m_\varrho^3}{M_P^2}\,.
\end{equation}
The dilaton couples to gravitinos and gauge degrees of freedom with equal strength. Similar as in the previous example, we can neglect the entropy release from gravitino decays.

Our third UV example is M-theory compactified on a $G_2$ manifold. The size and the shape of the
manifold is controlled by moduli $T_i$. The following ansatz for the K\"ahler potential has been suggested\footnote{We neglect a factor of $\pi/2$ which is irrelevant for our discussion.}~\cite{Beasley:2002db,Acharya:2005ez}
\begin{equation}
 K = -\log\left[\prod\limits_i(\bar{T}_i+T_i)^{a_i}\right]\,,\qquad \sum\limits_i a_i =7\,.
\end{equation}
The volume of the manifold is given by $\mathcal{V}=\prod\limits_i(\re T_i)^{a_i/3}$. It has been shown that all moduli can be stabilized by 2 or more gaugino condensates~\cite{Acharya:2007rc}. Hidden sector quarks charged under the confining gauge groups form meson fields $\phi_i$ which occur in the low energy theory. A simplified two-field description, which captures some of the main features, contains the modulus $T$ and the meson field $\phi$ (see~\cite{Kane:2019nod}). The K\"ahler potential reads
\begin{equation}
  K = -7 \log(\bar{T}+T) + \bar{\phi}\phi\,,
\end{equation}
and the gauge kinetic function $f=T$. Supersymmetry is dominantly broken by the meson. The cosmologically relevant light modulus field is $|\phi|$ with  $m_\phi \simeq 2\,m_{3/2}$.\footnote{The case with additional light moduli has been discussed in~\cite{Acharya:2008bk}.} In order, to reproduce the observed mass of the light Higgs boson $m_{3/2}\simeq 30-80\tev$ is required~\cite{Acharya:2008zi,Ellis:2014kla}. 
While $G_T$ is suppressed in the vacuum, the higher derivative $G_{TT}=\mathcal{O}(1)$. The mass of the heavier modulus $T$ is, hence, enhanced by a factor $K^{\bar{T}T}\sim\alpha_{\text{GUT}}^{-2}$ compared to the gravitino mass. For viable choices of the hidden sector gauge groups, one obtains $m_T\simeq (100-500)\times m_{3/2}$. The meson decay rate is dominated by contributions containing the derivative
\begin{equation}\label{eq:fderivative}
  \partial_\phi F^{\bar{T}} \simeq e^{G/2} K^{\bar{T}T} G_{T\phi} \simeq \sqrt{m_{3/2} m_T K^{\bar{T}T}}\,.
\end{equation}
In the last step, we employed the relation $G_{T\phi}\sim \sqrt{G_{TT}}$ which follows from eq.\ 19 in~\cite{Kane:2019nod}. The visible sector K\"ahler potential is expected to scale with the inverse volume $\mathcal{V}^{-1}\propto (\bar{T}+T)^{7/3}$~\cite{Acharya:2008hi} (while we assume it to be $\phi$-independent). We will, therefore, estimate $\partial_{T}Z \sim Z/T$, $\partial_{T}\hat Z \sim \hat Z/T$. The meson decay rates obtained from~\eqref{eq:Gammagg},~\eqref{eq:Gammagaga},~\eqref{eq:GammaHH} and~\eqref{eq:Gammahihi} are
\begin{equation}\label{eq:mesondecay}
\Gamma_{\tilde{\lambda}\tilde{\lambda}} \simeq
  \frac{N_{g}\,m_\phi^2\,m_T }{448\pi M_{\text{P}}^2}\,,\qquad
  \Gamma_{HH} \simeq \left(\frac{Z^2}{{\hat Z} ^2}+\frac{Z^4}{{\hat Z} ^4}\right)\frac{m_\phi^2\, m_T }{28\pi\,M_P^2}\,,\qquad
  \Gamma_{\tilde{h}\tilde{h}} \simeq \frac{Z^2}{{\hat Z} ^2}\frac{m_\phi^2\, m_T }{28\pi\,M_P^2}\,,
\end{equation}
while $\Gamma_{gg}$ and $\Gamma_{\tilde{f}\tilde{f}}$ are smaller. Notice that the meson decay rate is enhanced by a factor $m_T/m_\phi$ compared to the naive expectation. The size of $Z/\hat Z$ cannot be predicted from the UV theory. As motivated in~\cite{Acharya:2011te}, we will assume $\mu< m_{3/2}$ which implies $Z/\hat Z\lesssim 1$. In realistic $G_2$ compactifications, the gauge kinetic function depends on all moduli of the theory~\cite{Acharya:2007rc}. Furthermore, there is more freedom in the K\"ahler metric of $T$ (whose role is played by a linear combination of several moduli in the full theory). The decay rates derived above should, therefore, be considered to be correct within a factor of a few.\footnote{In order to be specific, we assume an additional uncertainty by a factor of 3 on the decay rates in~\eqref{eq:mesondecay}.}

In the M-theory model, gravitino final states are kinematically inaccessible in the meson decay. However, there appears a (nearly) massless axion which is dominantly the meson phase. For the canonical K\"ahler potential, the decay rate to axions vanishes (cf.~\eqref{eq:axionrate}). While higher order terms in the K\"ahler potential can open the axion channel, this decay mode remains suppressed compared to gauge and Higgs final states. This is because it does not receive the enhancement factor~\eqref{eq:fderivative}.

We summarize the results of the considered UV theories in table~\ref{tab:summary}. Decay rates are expressed in terms of the coefficients $c_{ii}$ which are defined by $\Gamma_{ii} = c_{ii}\, m_\varrho^3/M_P^2$
for the modulus $\varrho=S\,,T\,,\phi$. We included the uncertainties stated in the text. The table also contains the expected modulus abundance prior to decay as obtained from~\eqref{eq:yield}. Despite major differences in modulus decay patterns, $Y_\varrho$ agrees within one order of magnitude for the three theories considered here. In order to explain the observed baryon asymmetry of the universe, about $10^{-3}-10^{-4}$ baryons must be generated per modulus decay.

\begin{table}[htp]
\begin{center}
\begin{tabular}{|cccccccc|}
\hline 
\rowcolor{light-gray}&&&&&&&\\[-3mm]
\rowcolor{light-gray} model & $m_\varrho$ & $c_{gg}$ & $c_{\tilde{\lambda}\tilde{\lambda}}$ & $c_{HH}$ & $c_{\tilde{h}\tilde{h}}$ &  $c_{\tilde{\Psi}\tilde{\Psi}}$& $Y_\varrho$\\[1mm]
\hline
&&&&&&&\\[-3mm]
type IIb & $\!(1000-8000)\tev\!$ & $\!0.04\!$ & $\!0.04\!$ & $\!\lesssim 0.01\!$ & $\!\lesssim 0.01\!$ & $\!0.03\!$ & $\!(2-5)\times 10^{-7}\!$\\[1mm]
heterotic & $\!(2000-6000)\tev\!$ & $0.1$ & $0.1$ & $-$ & $-$ & $0.01$ & $(3-5)\times 10^{-7}$ \\[1mm]
M-theory & $(60-160)\tev$ & $-$ & $\!0.1-6\!$ & $\lesssim 20$ & $\lesssim 10$  & $-$ & $(0.4-10)\times 10^{-7}$\\ \hline
\end{tabular}
\end{center}
\caption{Modulus mass, decay coefficients and cosmological abundance (prior to decay) in different UV theories. Subleading decay channels are neglected.}
\label{tab:summary}
\end{table}

\section{Baryogenesis Mechanism}\label{sec:baryogenesis}

Since the modulus decay occurs out of thermal equilibrium, it automatically fulfills one of the Sakharov conditions~\cite{Sakharov:1967dj} required for successful baryogenesis. In addition, baryon number and CP violation is required to occur in the decay chain. Since moduli typically have a large (direct or indirect) branching ratio to gluinos, a natural possibility is to create baryons from intermediate gluinos. Baryon number is broken in the presence of the R-parity violating operator
\begin{equation}       
  W_{\text{RPV}}= \lambda^{\prime\prime}_{ijk} U^c_i D^c_j D^c_k\,.
\end{equation}
For simplicity, we will consider the case, where only $\lambda^{\prime\prime}_{323}=-\lambda^{\prime\prime}_{332}$ is non-vanishing and denote this coupling by $\lambda$ in the following. While strong experimental constraints on some combinations of R-parity violating couplings arise, $\lambda^{\prime\prime}_{323}$ itself can be large (see e.g.~\cite{Chemtob:2004xr}).

\begin{figure}[h]
\begin{center}   
 \includegraphics[width=10cm]{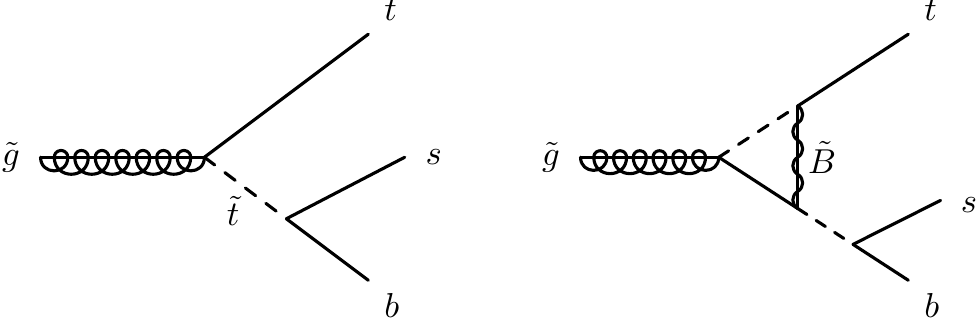}\hspace{6mm}
\end{center}
\caption{Baryon number violating gluino decays. The interference between tree-level and loop diagrams generates a baryon asymmetry.}
\label{fig:feynman_diagrams}
\end{figure}

The relevant baryon number violating decay is $\tilde{g} \rightarrow t\,s\,b\;(\bar{t}\,\bar{s}\,\bar{b})$. A CP asymmetry arises from the phase difference between gluino and bino (or wino) mass which is generically present in the MSSM.\footnote{In the presence of flavor violation, phases in the squark mixing matrix or the R-parity violating couplings can also contribute to the CP asymmetry~\cite{Cui:2013bta,Arcadi:2013jza,Arcadi:2015ffa}.} The latter shows up in the interference of the tree and loop diagrams shown in figure~\ref{fig:feynman_diagrams}.\footnote{Alternative baryogenesis mechanisms employing hadronic R-parity violation have been suggested in~\cite{Dimopoulos:1987rk,Cline:1990bw,Kohri:2009ka,Sorbello:2013xwa}. These, however, require a superpartner spectrum which is different from the expected spectrum in the UV theories discussed in this work.}
Previously, baryogenesis by gaugino decay has been considered for a thermal history of the universe~\cite{Cui:2013bta,Arcadi:2013jza,Arcadi:2015ffa} (see also~\cite{Baldes:2014rda}). However, in this case, the gaugino decays at a temperature, at which washout reactions are typically still active. They tend to erase the previously produced baryon asymmetry. On the other hand, the wash-out problem completely disappears if baryons are created from modulus decay. This is because the decay temperature is so low that wash-out processes are extremely suppressed. 

The final baryon abundance from modulus decay can then be written in the form
\begin{equation}
  Y_b = Y_{\tilde{g}} \; \epsilon_{\text{CP}}\,,
\end{equation}
where $\epsilon_{\text{CP}}$ denotes the CP asymmetry and $Y_{\tilde{g}}$ the gluino abundance (prior to decay). The latter is determined by the modulus abundance times the (direct or indirect) branching ratio into gluinos,\footnote{We neglect a subdominant contribution from intermediate gravitinos.}
\begin{equation}
  Y_{\tilde{g}} = \frac{4}{3}\text{Br}(\varrho\rightarrow\tilde{\lambda}\tilde{\lambda}) 
  \, Y_\varrho\,.
\end{equation}
For the UV theories discussed in the previous section we obtain
\begin{equation}\label{eq:gluinoabundance}
  Y_{\tilde{g}} = \begin{cases} (1-4)\times 10^{-7} & \text{(type IIb),} \\(2-4)\times 10^{-7} & \text{(heterotic),}\\  (0.03-6)\times 10^{-7} & \text{(M-theory).} \end{cases}
\end{equation}
The CP asymmetry is determined as
\begin{equation}\label{eq:cp}
  \epsilon_{\text{CP}} = \frac{\Gamma(\tilde{\lambda}_1\rightarrow t\, s\, b-\bar{t}\, \bar{s}\, \bar{b})}{\Gamma(\tilde{\lambda}_1\rightarrow t\, s\, b+\bar{t}\, \bar{s}\, \bar{b})+\Gamma(\tilde{\lambda}_1\rightarrow \tilde{\lambda}_2 \:t\,\bar{t} )}\;.
\end{equation}
Notice that besides the two diagrams of figure~\ref{fig:feynman_diagrams}, the R-parity conserving decay $\tilde{\lambda}_1\rightarrow \tilde{\lambda}_2 \:t\,\bar{t} $ contributes to the total decay rate. 

In the MSSM implementation we considered so far, $\tilde{\lambda}_1$ is identified with the gluino and $\tilde{\lambda}_2$ can be the bino or the wino. We will later also discuss baryogenesis through a hidden sector gaugino. Therefore, we present the decay rates for general $\tilde{\lambda}_{1,2}$. We assume that squarks are significantly heavier than gauginos as motivated by the UV models we considered. For simplicity, we completely decouple all sfermions other than the lightest stop $\widetilde{t}_1$ which we take to be an arbitrary linear combination
\begin{equation}
  \tilde{t}_1 = c_t\, \tilde{t}_R + s_t \,\tilde{t}_L\,.
\end{equation}
For the sake of a compact notation, we introduced $c_t=\cos\theta_{\tilde{t}}$, $s_t=\sin\theta_{\tilde{t}}$. The simpler case without left-right mixing has been discussed in~\cite{Cui:2013bta,Arcadi:2013jza,Arcadi:2015ffa,Pierce:2019ozl}. We observe, however, that non-trivial $\theta_{\tilde{t}}$ can significantly affect the baryon asymmetry. In particular, the diagram with a wino in the loop only contributes in the presence of left-right mixing. This is because the wino couples only to left-handed states, while the R-parity violation coupling operates on right-handed states. We also note that significant left-right mixing naturally occurs in the stop sector due to the large top Yukawa coupling. Neglecting other sfermion states is justified since $\tilde{t}_1$ usually comes out as the lightest squark due to the renormalization group running. Hence, $\tilde{t}_1$-mediated processes typically dominate the rates. In any case, the inclusion of further squarks would increase CP violating and CP conserving decay modes in a similar fashion such that $\epsilon_{\text{CP}}$ is qualitatively not affected. With the mentioned assumptions we find
\begin{align}
&\Gamma(\tilde{\lambda}_1\rightarrow t\, s\, b+\bar{t}\, \bar{s}\, \bar{b})=\frac{C_a\,\lambda ^2 m_{\tilde{\lambda}_1}^5 }{768 \pi ^3 m_{\tilde{t}}^4} \left(\kIL^2\,s_t^2 \,c_t^2+\kIR^2 \, c_t^4\right)\,, \label{eq:tree1}\\
&\Gamma(\tilde{\lambda}_1\rightarrow \tilde{\lambda}_2 \:t\,\bar{t} ) =\frac{C_b\,m_{\tilde{\lambda}_1}^5}{512 \pi^3\, m_{\tilde{t}}^4} \left[
\frac{8}{3}\frac{m_{\tilde{\lambda}_2}}{m_{\tilde{\lambda}_1}}\,
      c_t^2 
       s_t^2 \,\kIL \,\kIR \,\kIIL \,
    \kIIR \:f_3\!\left(\tfrac{m_{\tilde{\lambda}_2}^2}{m_{\tilde{\lambda}_1}^2}\right)\right.\nonumber\\
    & \qquad\qquad\qquad\;\; +
\left(
     \kIR^2 \,\kIIR^2\, c_t^4 + 
       \kIL^2 \,\kIIL^2 \,s_t^4 + 
        \frac{2}{3} \,c_t^2 
        s_t^2 \,\left(\kIL^2 \,\kIIR^2 + 
                 \kIR^2 \,\kIIL^2\right)\right)\left.
    f_2\!\left(\tfrac{m_{\tilde{\lambda}_2}^2}{m_{\tilde{\lambda}_1}^2}\right)
    \right]\,,\label{eq:tree2}\\
&\Gamma(\tilde{\lambda}_1\rightarrow t\, s\, b-\bar{t}\, \bar{s}\, \bar{b})=\frac{C_c\,\lambda ^2 m_{\tilde{\lambda}_1}^5 }{768 \pi ^3 m_{\tilde{t}}^4} \left[
\frac{m_{\tilde{\lambda}_1}^2}{20 \pi\,  m_{\tilde{t}}^2}\,\kIR \, \kIL \, \kIIR\,\kIIL\,c_t^4\,s_t^2\;f_1\!\left(\tfrac{m_{\tilde{\lambda}_2}^2}{m_{\tilde{\lambda}_1}^2}\right)\right.
\nonumber\\&\qquad\qquad\qquad\qquad\qquad\quad
\left.
+\frac{m_{\tilde{\lambda}_1} m_{\tilde{\lambda}_2}}{16 \pi\,  m_{\tilde{t}}^2}  
\left( \kIR^2 \, \kIIR^2\, c_t^6 +   \kIL^2  \,\kIIL^2\, c_t^2\, s_t^4\right)
f_2\!\left(\tfrac{m_{\tilde{\lambda}_2}^2}{m_{\tilde{\lambda}_1}^2}\right)
\right] \,\sin\left(2\phi_{12}\right)\,,\label{eq:loop}
\end{align}
where we introduced
\begin{align}
  f_1(x) &= (1 - x)^5\,,\qquad
  f_2(x) = 1 - 8 x + 8 x^3 - x^4 - 12 x^2 \log x\,,\nonumber\\
  f_3(x) &= 1 + 9 x - 9 x^2 - x^3 + 6 (x+x^2) \log x\,,
\end{align}
and abbreviated $m_{\tilde{t}_1}$ by $m_{\tilde{t}}$. Notice that a non-vanishing CP asymmetry requires $m_{\tilde{\lambda}_1}>m_{\tilde{\lambda}_2}$ consistent with the Nanopoulos-Weinberg theorem~\cite{Nanopoulos:1979gx}. The coefficients $C_{a,b,c}$ denote color factors, while the $\kappa$ are the left- and right handed gaugino couplings. Finally $\phi_{12}=\phi_{\tilde{\lambda}_2}-\phi_{\tilde{\lambda}_1}$ denotes the phase difference between the decaying gaugino and the gaugino running in the loop. Parameter values for the MSSM baryogenesis implementation $ \tilde {\lambda }_1=\tilde{g}$, $\tilde{\lambda}_2=\tilde{B},\tilde{W}$ are listed in table~\ref{tab:baryocouplings}.

\begin{table}[htp]
\begin{center}
\begin{tabular}{|ccccccccc|}
\hline 
\rowcolor{light-gray}&&&&&&&&\\[-3mm]
\rowcolor{light-gray} $\tilde{\lambda}_1$ & $\tilde{\lambda}_2$ & $C _a$ & $C_b$ & $C_c$ & $\kIR$ & $\kIL$ & $\kIIR$ & $\kIIL$\\[1mm]
\hline
&&&&&&&&\\[-3mm]
$\tilde{g}$ & $\tilde{B}$ & $1$ & $\frac{1}{2}$ & $1$ & $\sqrt{2}\,g_3$ & $-\sqrt{2}\,g_3$ & $\frac{2\sqrt{2}}{3} g_1 $ & $-\frac{\sqrt{2}}{6} g_1 $\\[1mm]
$\tilde{g}$ & $\tilde{W}$ & $1$ & $\frac{1}{2}$ & $1$ & $\sqrt{2}\,g_3$ & $-\sqrt{2}\,g_3$ & $-$ & $-\frac{1}{\sqrt{2}}g_2$\\[1mm]
$\tilde{B}^\prime$ & $\tilde{g}$ & $6$ & $4$ & $8$ & $\frac{2\sqrt{2}}{3} \varepsilon\,g_1 $ & $-\frac{\sqrt{2}}{6} \varepsilon\,g_1 $ & $\sqrt{2}\,g_3$ & $-\sqrt{2}\,g_3$\\ \hline
\end{tabular}
\end{center}
\caption{Color factors and couplings entering~\eqref{eq:tree1},~\eqref{eq:tree2} and~\eqref{eq:loop}. The first two rows refer to the MSSM implementation of baryogenesis. In this case, the decay of a gluino with a bino or wino in the loop is considered. The last row refers to the hidden sector implementation with a hidden bino decaying through a gluino loop.}
\label{tab:baryocouplings}
\end{table}

In order to assess the CP asymmetry, it is instructive to consider the limit of a purely right-handed $\tilde{t}_1$. In this case, only the bino contributes and we obtain the CP asymmetry
\begin{align}
  \epsilon_{\text{CP}} &= \frac{g_1^2\,m_{\tilde{g}}\,m_{\tilde{B}}\,f_2\!\left(\tfrac{m_{\tilde{B}}^2}{m_{\tilde{g}}^2}\right)}{18\pi\,m_{\tilde{t}}^2}\,\left(1+\frac{g_1^2}{6\lambda^2} f_2\!\left(\tfrac{m_{\tilde{B}}^2}{m_{\tilde{g}}^2}\right)\right)^{-1}\,\sin\left(2\phi_{12}\right)\nonumber\\
  &\!\!\!\overset{\lambda\gg g_1}{\simeq}\; 1.1\cdot 10^{-4} \,\left(\frac{m_{\tilde{g}}}{5\tev}\right) \left(\frac{m_{\tilde{B}}}{1\tev}\right) \left(\frac{10\tev}{m_{\tilde{t}}}\right)^2 f_2\!\left(\tfrac{m_{\tilde{B}}^2}{m_{\tilde{g}}^2}\right)\,\sin\left(2\phi_{12}\right)\,.
\end{align}
The kinematical factor $f_2$ takes values between 0 and 1. For the benchmark case $m_{\tilde{B}}/m_{\tilde{g}}=1/5$, one finds $f_2\simeq 0.7$. This suggests $\epsilon_{\text{CP}}\lesssim 10^{-4}\,\sin(2\phi_{12})$ in the case of a purely right-handed stop. Given that $Y_{\tilde{g}}< 10^{-6}$ in the UV models we described (cf.~\eqref{eq:gluinoabundance}), a large enough baryon asymmetry can, hence, only be generated for large relative gaugino phase. In the described UV models, some suppression of the gaugino phase is, however, expected~\cite{Kane:2009kv}. Typically $\phi_{12}\simeq 0$ at the UV scale. It gets generated radiatively or by threshold effects (see e.g.~\cite{Olive:2005ru}). 

\begin{figure}[h]
\begin{center}   
 \includegraphics[width=11.5cm]{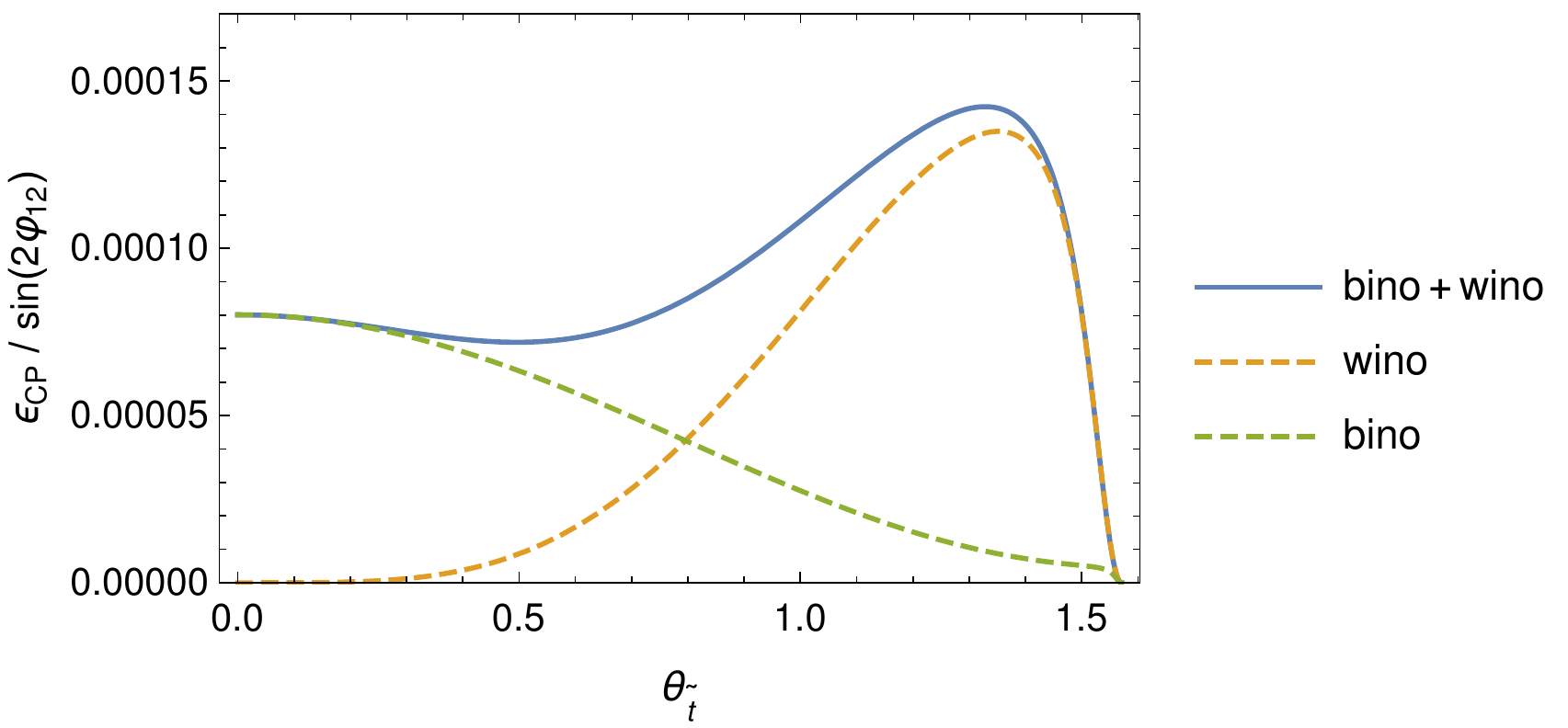}
\end{center}
\caption{CP asymmetry as a function of the stop mixing angle. We set $m_{\tilde{t}}=10\tev$, $m_{\tilde{g}}=5\tev$, $m_{\tilde{W}}=1.5\tev$, $m_{\tilde{B}}=1\tev$ and $\lambda=2$.}
\label{fig:angle}
\end{figure}

Let us now investigate, whether the CP asymmetry gets enhanced in the presence of stop left-right mixing. In figure~\ref{fig:angle} we plot the CP asymmetry as a function of $\theta_{\tilde{t}}$. For non-trivial mixing, the wino loop contributes. We have assumed that wino and bino have a universal phase which appears reasonable since renormalization group running mostly affects the gluino phase. Masses are stated in the figure caption. We observe that left-right mixing can increase the CP asymmetry by a factor $\sim 2$. Notice that the wino contribution vanishes in the limit of a purely right- or left-handed state.

If we choose the optimal $\theta_{\tilde{t}}$ and vary the masses, we find $\epsilon_{\text{CP}}\lesssim 3\cdot 10^{-4}$. For the maximal value of $Y_{\tilde{g}}\simeq 6\cdot 10^{-7}$ motivated by the UV theories, we see that a sufficiently large baryon asymmetry requires
\begin{equation}
  \theta_{12}\gtrsim 10^\circ\,.
\end{equation}
A value of this size appears still reasonable in the light of some phase suppression (expected from the UV argument). 
While strong experimental constraints on gaugino phases arise from electric dipole moments, these depend on the origin of the phases. Large (even maximal) $\phi_{12}$ is e.g.\ observationally acceptable if it originates from the gluino, while all electroweak parameters are real. This was explicitly verified for the superpartner spectrum we consider with the code SPheno~4.0.3~\cite{Porod:2003um,Porod:2011nf}. Nevertheless, electric dipole moment measurements offer great opportunities to probe the baryogenesis mechanism within a concrete UV theory, in which the pattern of phases is predicted.

We conclude that the MSSM baryogenesis implementation could work. It may appear uncomfortable that one has to push uncertainties a bit in order to maximize the baryon asymmetry. On the other hand, moduli stabilization mechanisms beyond those considered in this work could potentially give rise to a slightly higher modulus abundance and simplify the generation of the observed baryon asymmetry. 

Nevertheless, we should ask, whether the baryon asymmetry can be enhanced within a slight modification of the mechanism.
A very natural possibility is to include a hidden sector of particles. In the next section, we will entertain the possibility that the baryon asymmetry is generated by the decay of a hidden sector gaugino rather than the gluino~\cite{Pierce:2019ozl}.

\section{Hidden Sector Implementation}

Realistic string/ M-theory compactifications predict a large number of new states beyond the MSSM (see e.g.~\cite{Acharya:2017kfi}). Many of those are uncharged under the Standard Model gauge groups and only interact with visible matter through suppressed portal couplings. While the existence of such hidden sector particles is difficult to probe at accelerator experiments, it may significantly affect cosmology. In particular, when the universe was reheated by modulus decay, visible and hidden matter should have been produced in similar abundance. This is because the modulus couplings are gravitational in nature and insensitive to gauge charges. Intriguingly, the decay of the hidden sector particles may have generated the baryon asymmetry as we describe in the following.

As a minimal realization of a hidden sector, we consider a $U(1)^\prime$ gauge theory which kinetically mixes with the hypercharge $U(1)_Y$~\cite{Holdom:1985ag}. We assume that a Higgs mechanism (including two hidden Higgs superfields as required for anomaly cancellation) operates in the hidden sector. The hidden sector mass spectrum could resemble that of the electroweak gauge bosons and neutralinos since gravity mediation naturally induces soft masses of comparable size. We take the hidden bino $B^\prime$ to be the lightest hidden sector fermion which must decay into MSSM fields through the (small) portal coupling.

\begin{figure}[h]
\begin{center}   
 \includegraphics[width=10cm]{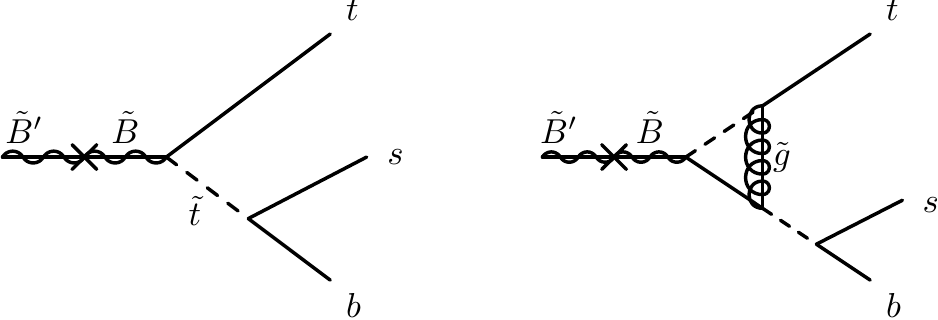}\hspace{6mm}
\end{center}
\caption{Baryon number violating hidden sector bino decays. The interference between tree-level and loop diagrams generates a baryon asymmetry.}
\label{fig:feynman_diagrams2}
\end{figure}

Given that at least one of the MSSM gauginos is lighter than $B^\prime$, a baryon asymmetry is generated by the process $\tilde{B}^\prime \rightarrow t\,s\,b\;(\bar{t}\,\bar{s}\,\bar{b})$. The CP asymmetry arises from the phase difference between the hidden sector bino and one or several MSSM gauginos running in the loop (see figure~\ref{fig:feynman_diagrams2}). The case where only the bino contributes has been studied in~\cite{Pierce:2019ozl}. If $m_{\tilde{B}^\prime}>m_{\tilde{g}}$, the baryon asymmetry is instead typically dominated by the gluino loop. This is the case we will focus on.

\noindent
The baryon asymmetry can be written in the familiar form
\begin{equation}
  Y_b = Y_{\tilde{B}^\prime} \; \epsilon_{\text{CP}}\,,
\end{equation}
where $Y_{\tilde{B}^\prime}$ denotes the intermediate hidden bino abundance resulting from modulus decay. The latter is more model-dependent compared to the gluino abundance discussed in section~\ref{sec:baryogenesis}. Depending on the geometric localization of the hidden sector, the $U(1)^\prime$ gauge kinetic function may e.g.\ carry a different moduli-dependence compared to the visible sector gauge groups. Therefore, we will limit the discussion to a simple example. The light modulus decays into higgsinos with an effective coupling up to $c_{\tilde{h}\tilde{h}}\sim 10$ (see M-theory case in table~\ref{tab:summary}). A similar coupling is expected for the hidden sector higgsinos which subsequently decay into hidden binos. We estimate that the resulting hidden bino abundance from this channel,
\begin{equation}
  Y_{\tilde{B}^\prime} =  2\,\text{Br}(\varrho\rightarrow\tilde{h}^\prime\tilde{h}^\prime) \, Y_\varrho\,,
\end{equation}
can be as large as $Y_{\tilde{B}^\prime}\sim 10^{-6}$. Additional channels, including the direct decay of the modulus into hidden binos, may further increase $Y_{\tilde{B}^\prime}$, but unlikely by more than an $\mathcal{O}(1)$ factor.

The CP asymmetry is again determined by~\eqref{eq:cp}. This time $\tilde{\lambda}_1$ is identified with the hidden bino and $\tilde{\lambda}_2$ with the gluino. The decay rates for the tree and loop processes are those of~\eqref{eq:tree1}$\,$--$\,$\eqref{eq:loop}, with the coupling and color factors from table~\ref{tab:baryocouplings} (last line). The hidden bino-visible bino mixing parameter is denoted by $\varepsilon$. In order to estimate the CP asymmetry numerically, we first consider the limit of a purely right-handed $\tilde{t}_1$. For this case, we find
\begin{align}
  \epsilon_{\text{CP}} &= \frac{g_3^2\,m_{\tilde{g}}\,m_{\tilde{B}^\prime}\,f_2\!\left(\tfrac{m_{\tilde{g}}^2}{m_{\tilde{B}^\prime}^2}\right)}{6\pi\,m_{\tilde{t}}^2}\,\left(1+\frac{g_3^2}{2\lambda^2} f_2\!\left(\tfrac{m_{\tilde{g}}^2}{m_{\tilde{B}^\prime}^2}\right)\right)^{-1}\,\sin\left(2\phi_{12}\right)\nonumber\\
  &\!\!\!\overset{\lambda\gg g_3}{\simeq}\; 1.2\cdot 10^{-2} \,\left(\frac{m_{\tilde{B}^\prime}}{6\tev}\right) \left(\frac{m_{\tilde{g}}}{3\tev}\right) \left(\frac{10\tev}{m_{\tilde{t}}}\right)^2 f_2\!\left(\tfrac{m_{\tilde{g}}^2}{m_{\tilde{B}^\prime}^2}\right)\,\sin\left(2\phi_{12}\right)\,,
\end{align}
where $f_2\simeq 0.2$ for a mass ratio $m_{\tilde{g}}/m_{\tilde{B}^\prime}=1/2$. Notice that the CP asymmetry is significantly larger compared to the visible sector implementation described in section~\ref{sec:baryogenesis}. The increase essentially comes from the larger gauge coupling of the gluino in the loop compared to the electroweak gauginos.

For the case of hidden sector baryogenesis, left-right mixing in the stop sector is of minor interest. Since the gluino couples equally strong to left- and right-handed fields, the relative strength of the baryon number violating tree and loop processes in figure~\ref{fig:feynman_diagrams2} is rather insensitive to the stop mixing angle. The baryon number conserving decay $\tilde{B}^\prime \rightarrow \tilde{g} \:t\,\bar{t} $, however, gains importance for $\theta_{\tilde{t}}>0$ and slightly suppresses the CP asymmetry. We can, therefore, concentrate on the case of a right-handed $\tilde{t}_1$.

\begin{figure}[h]
\begin{center}   
  \includegraphics[width=8.5cm]{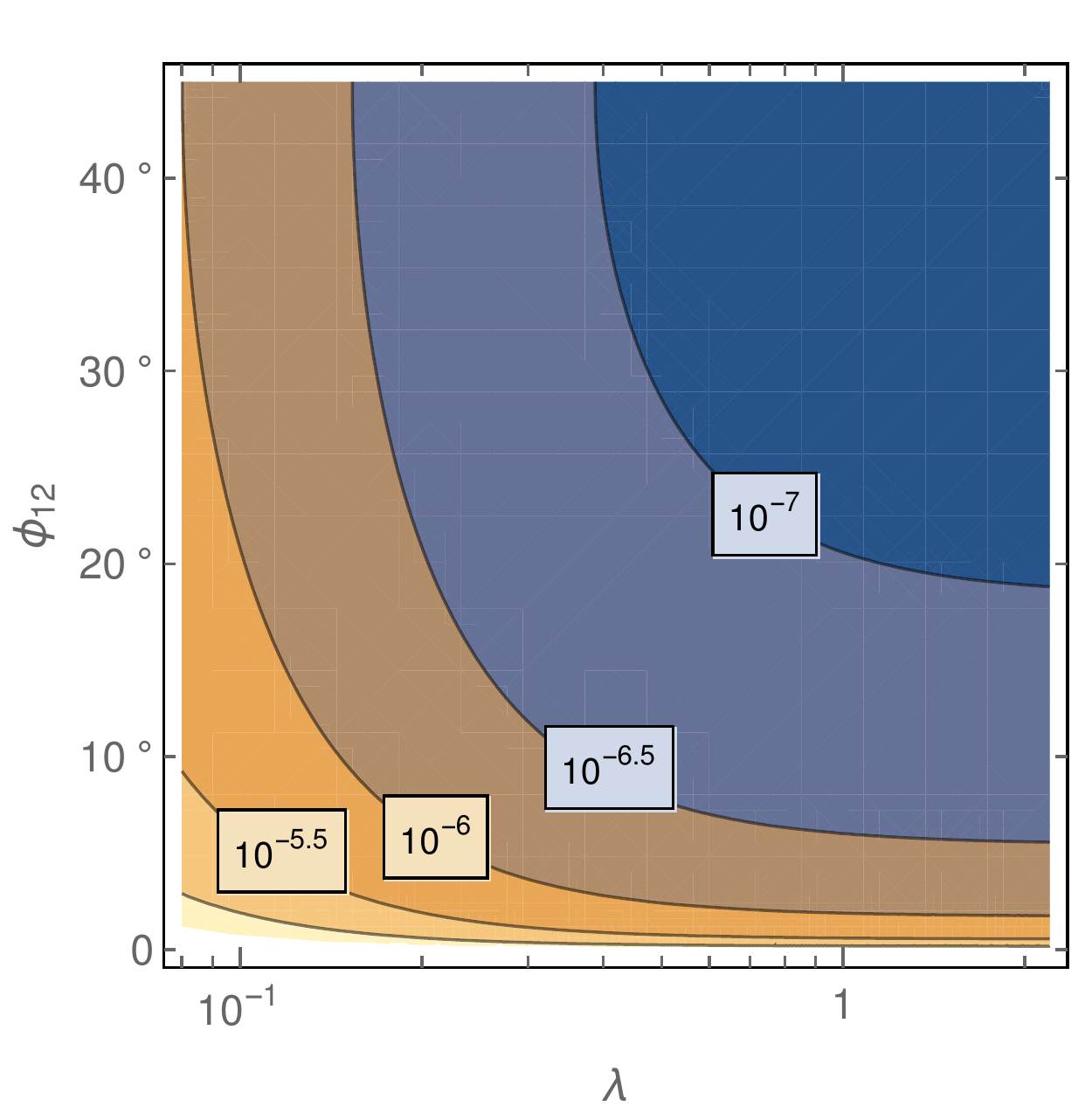}
\end{center}
\caption{Hidden bino abundance $Y_{\tilde{B}^\prime}$ required for successful baryogenesis as a function of the R-parity violating coupling $\lambda$ and the relative phase $\phi_{12}$ (in units of degrees) between hidden bino and gluino. The gaugino and stop masses were fixed to $m_{\tilde{B}^\prime}=6\tev$, $m_{\tilde{g}}=3\tev$ and $m_{\tilde{t}}=12\tev$. Values of at least $Y_{\tilde{B}^\prime}\sim 10^{-6}$ can be realized in well-motivated UV models.}
\label{fig:binoabundance}
\end{figure}

In figure~\ref{fig:binoabundance} we scanned the hidden bino abundance required for successful baryogenesis. For realistic $Y_{\tilde{B}^\prime}\sim 10^{-6}$, the observed baryon asymmetry is realized with moderate values of the phase and the R-parity violating coupling (e.g. $\phi_{12}\sim 5^\circ$ and $\lambda\sim 0.2$). The described hidden sector baryogenesis mechanism, hence, works very naturally in wide regions of the parameter space.

\section{Conclusion}

String/ M-theory often gives rise to a ``light'' modulus which alters the cosmological history. Its late decay at $T\simeq 10-100\mev$ releases large amounts of entropy and washes out any previously produced matter asymmetry. While this is often considered as (part of) the cosmological moduli problem, we argued that the decay of the modulus can successfully generate the baryon asymmetry.

In the first step, we explicitly calculated the modulus abundance and decay pattern. This was done in well-motivated UV theories including flux compactifications of type IIb string theory, orbifold compactifications of the heterotic string as well as compactifications of M-theory on $G_2$ manifolds. Despite major differences in the underlying UV theories we found a rather universal prediction of $Y_\varrho=10^{-7}-10^{-6}$ for the modulus abundance prior to decay. This implies that about $10^{-4}-10^{-3}$ baryons must effectively be created per modulus -- a value not much smaller than a loop factor.

Since we generically found a large branching ratio of the modulus into gauginos, we reasoned that modulus decay through an intermediate gaugino is the prime candidate for realizing baryogenesis. CP violation naturally occurs via the relative phase in the gaugino mass matrix. The latter shows up in the interference between the tree- and loop-level gaugino decay diagrams. For the required baryon number breaking we considered the R-parity violating operator $W\supset t^c b^c s^c$.

We then determined the baryon asymmetry including only the fields of the MSSM. In this case, the dominant contribution arises from the decay of the gluino with a bino (wino) running in the loop. We investigated the impact of left-right mixing in the stop sector and found that it can enhance the resulting baryon asymmetry by up to a factor $\sim 2$. While the observed matter abundance can be realized within this scheme, a relatively large phase of $\phi_{12} \gtrsim 10^\circ$ between the gluino and the bino (wino) is required. We then showed that the baryon asymmetry is significantly enhanced if the MSSM is extended by a hidden sector bino field which kinematically mixes with the visible bino. In the hidden sector implementation, successful baryogenesis occurs for moderate gaugino phase $\phi_{12}\gtrsim 1^\circ$.

Future measurements of electric dipole moments offer exciting prospects to test the implementation of baryogenesis we describe. We expect this mechanism to at least qualitatively provide baryogenesis in several corners of compactified string- and M-theory.

\section*{Acknowledgments}
We would like to thank Aaron Pierce and Bibhushan Shakya for very helpful discussions. MW acknowledges support by the Vetenskapsr\r{a}det (Swedish Research Council) through contract No. 638-2013-8993 and the Oskar Klein Centre for Cosmoparticle Physics and the LCTP at the University of Michigan, and both of us from DoE grant DE-SC0007859.

\bibliography{moduli}

\begin{thebibliography}{10}

\bibitem{Coughlan:1983ci}
G.D. Coughlan et~al.,
\newblock Phys. Lett. 131B (1983) 59.

\bibitem{Kitano:2008tk}
R. Kitano, H. Murayama and M. Ratz,
\newblock Phys. Lett. B669 (2008) 145, 0807.4313.

\bibitem{Allahverdi:2010im}
R. Allahverdi, B. Dutta and K. Sinha,
\newblock Phys. Rev. D82 (2010) 035004, 1005.2804.

\bibitem{Allahverdi:2010rh}
R. Allahverdi, B. Dutta and K. Sinha,
\newblock Phys. Rev. D83 (2011) 083502, 1011.1286.

\bibitem{Ishiwata:2013waa}
K. Ishiwata, K.S. Jeong and F. Takahashi,
\newblock JHEP 02 (2014) 062, 1312.0954.

\bibitem{Dhuria:2015xua}
M. Dhuria, C. Hati and U. Sarkar,
\newblock Phys. Lett. B756 (2016) 376, 1508.04144.

\bibitem{Cui:2013bta}
Y. Cui,
\newblock JHEP 12 (2013) 067, 1309.2952.

\bibitem{Arcadi:2013jza}
G. Arcadi, L. Covi and M. Nardecchia,
\newblock Phys. Rev. D89 (2014) 095020, 1312.5703.

\bibitem{Arcadi:2015ffa}
G. Arcadi, L. Covi and M. Nardecchia,
\newblock Phys. Rev. D92 (2015) 115006, 1507.05584.

\bibitem{Pierce:2019ozl}
A. Pierce and B. Shakya,
\newblock JHEP 06 (2019) 096, 1901.05493.

\bibitem{Buchmuller:2004xr}
W. Buchmuller et~al.,
\newblock Nucl. Phys. B699 (2004) 292, hep-th/0404168.

\bibitem{deSalas:2015glj}
P.F. de~Salas et~al.,
\newblock Phys. Rev. D92 (2015) 123534, 1511.00672.

\bibitem{Hasegawa:2019jsa}
T. Hasegawa et~al.,
\newblock (2019), 1908.10189.

\bibitem{Brignole:1997dp}
A. Brignole, L.E. Ibanez and C. Munoz,
\newblock Adv. Ser. Direct. High Energy Phys. 18 (1998) 125, hep-ph/9707209.

\bibitem{Nakamura:2006uc}
S. Nakamura and M. Yamaguchi,
\newblock Phys. Lett. B638 (2006) 389, hep-ph/0602081.

\bibitem{Endo:2006zj}
M. Endo, K. Hamaguchi and F. Takahashi,
\newblock Phys. Rev. Lett. 96 (2006) 211301, hep-ph/0602061.

\bibitem{Higaki:2013lra}
T. Higaki, K. Nakayama and F. Takahashi,
\newblock JHEP 07 (2013) 005, 1304.7987.

\bibitem{Kachru:2003aw}
S. Kachru et~al.,
\newblock Phys. Rev. D68 (2003) 046005, hep-th/0301240.

\bibitem{Cicoli:2012aq}
M. Cicoli, J.P. Conlon and F. Quevedo,
\newblock Phys. Rev. D87 (2013) 043520, 1208.3562.

\bibitem{Hebecker:2014gka}
A. Hebecker et~al.,
\newblock JHEP 09 (2014) 140, 1403.6810.

\bibitem{Giddings:2001yu}
S.B. Giddings, S. Kachru and J. Polchinski,
\newblock Phys. Rev. D66 (2002) 106006, hep-th/0105097.

\bibitem{Lebedev:2006qq}
O. Lebedev, H.P. Nilles and M. Ratz,
\newblock Phys. Lett. B636 (2006) 126, hep-th/0603047.

\bibitem{ORaifeartaigh:1975nky}
L. O'Raifeartaigh,
\newblock Nucl. Phys. B96 (1975) 331.

\bibitem{Kallosh:2006dv}
R. Kallosh and A.D. Linde,
\newblock JHEP 02 (2007) 002, hep-th/0611183.

\bibitem{Antoniadis:2014oya}
I. Antoniadis et~al.,
\newblock Phys. Lett. B733 (2014) 32, 1403.3269.

\bibitem{Kallosh:2014wsa}
R. Kallosh and T. Wrase,
\newblock JHEP 12 (2014) 117, 1411.1121.

\bibitem{Choi:2005ge}
K. Choi et~al.,
\newblock Nucl. Phys. B718 (2005) 113, hep-th/0503216.

\bibitem{Falkowski:2005ck}
A. Falkowski, O. Lebedev and Y. Mambrini,
\newblock JHEP 11 (2005) 034, hep-ph/0507110.

\bibitem{Moroi:1999zb}
T. Moroi and L. Randall,
\newblock Nucl. Phys. B570 (2000) 455, hep-ph/9906527.

\bibitem{Lowen:2008fm}
V. Lowen and H.P. Nilles,
\newblock Phys. Rev. D77 (2008) 106007, 0802.1137.

\bibitem{Lebedev:2006tr}
O. Lebedev et~al.,
\newblock Phys. Rev. Lett. 98 (2007) 181602, hep-th/0611203.

\bibitem{Krippendorf:2012ir}
S. Krippendorf et~al.,
\newblock Phys. Lett. B712 (2012) 87, 1201.4857.

\bibitem{Badziak:2012yg}
M. Badziak et~al.,
\newblock JHEP 03 (2013) 094, 1212.0854.

\bibitem{Beasley:2002db}
C. Beasley and E. Witten,
\newblock JHEP 07 (2002) 046, hep-th/0203061.

\bibitem{Acharya:2005ez}
B.S. Acharya, F. Denef and R. Valandro,
\newblock JHEP 06 (2005) 056, hep-th/0502060.

\bibitem{Acharya:2007rc}
B.S. Acharya et~al.,
\newblock Phys. Rev. D76 (2007) 126010, hep-th/0701034.

\bibitem{Kane:2019nod}
G. Kane and M.W. Winkler,
\newblock Phys. Rev. D100 (2019) 066005, 1902.02365.

\bibitem{Acharya:2008bk}
B.S. Acharya et~al.,
\newblock JHEP 06 (2008) 064, 0804.0863.

\bibitem{Acharya:2008zi}
B.S. Acharya et~al.,
\newblock Phys. Rev. D78 (2008) 065038, 0801.0478.

\bibitem{Ellis:2014kla}
S.A.R. Ellis, G.L. Kane and B. Zheng,
\newblock JHEP 07 (2015) 081, 1408.1961.

\bibitem{Acharya:2008hi}
B.S. Acharya and K. Bobkov,
\newblock JHEP 09 (2010) 001, 0810.3285.

\bibitem{Acharya:2011te}
B.S. Acharya et~al.,
\newblock JHEP 05 (2011) 033, 1102.0556.

\bibitem{Sakharov:1967dj}
A.D. Sakharov,
\newblock Pisma Zh. Eksp. Teor. Fiz. 5 (1967) 32,
\newblock [Usp. Fiz. Nauk161,no.5,61(1991)].

\bibitem{Chemtob:2004xr}
M. Chemtob,
\newblock Prog. Part. Nucl. Phys. 54 (2005) 71, hep-ph/0406029.

\bibitem{Dimopoulos:1987rk}
S. Dimopoulos and L.J. Hall,
\newblock Phys. Lett. B196 (1987) 135.

\bibitem{Cline:1990bw}
J.M. Cline and S. Raby,
\newblock Phys. Rev. D43 (1991) 1781.

\bibitem{Kohri:2009ka}
K. Kohri, A. Mazumdar and N. Sahu,
\newblock Phys. Rev. D80 (2009) 103504, 0905.1625.

\bibitem{Sorbello:2013xwa}
F. Rompineve,
\newblock JHEP 08 (2014) 014, 1310.0840.

\bibitem{Baldes:2014rda}
I. Baldes et~al.,
\newblock JCAP 1411 (2014) 041, 1410.0108.

\bibitem{Nanopoulos:1979gx}
D.V. Nanopoulos and S. Weinberg,
\newblock Phys. Rev. D20 (1979) 2484.

\bibitem{Kane:2009kv}
G. Kane, P. Kumar and J. Shao,
\newblock Phys. Rev. D82 (2010) 055005, 0905.2986.

\bibitem{Olive:2005ru}
K.A. Olive et~al.,
\newblock Phys. Rev. D72 (2005) 075001, hep-ph/0506106.

\bibitem{Porod:2003um}
W. Porod,
\newblock Comput. Phys. Commun. 153 (2003) 275, hep-ph/0301101.

\bibitem{Porod:2011nf}
W. Porod and F. Staub,
\newblock Comput. Phys. Commun. 183 (2012) 2458, 1104.1573.

\bibitem{Acharya:2017kfi}
B.S. Acharya et~al.,
\newblock JHEP 09 (2018) 130, 1707.04530.

\bibitem{Holdom:1985ag}
B. Holdom,
\newblock Phys. Lett. 166B (1986) 196.

\end{thebibliography}
\bibliographystyle{elsevier}
\end{document}